\documentstyle[twocolumn,aps]{revtex}

\newcommand{\be}{\begin{equation}}
\newcommand{\ee}{\end{equation}}
\newcommand{\bea}{\begin{eqnarray}}
\newcommand{\eea}{\end{eqnarray}}

\newcommand{\donedtbar}{\frac{\partial \bar{\psi_1}}{\partial \bar{t}}}
\newcommand{\dtwodtbar}{\frac{\partial \bar{\psi_2}}{\partial \bar{t}}}

\begin{document}
\author{Eric L. Bolda and Dan F. Walls}

\address{Department of Physics, University of Auckland, Private Bag 
92019,\\ 
Auckland, New Zealand.}

\title{Creation of vortices in a Bose-Einstein condensate by a Raman technique}

\date{\today}
\draft

\maketitle
  
\begin{abstract}
We propose a method for taking a Bose-Einstein condensate in the ground 
trap state simultaneously to a different atomic hyperfine state and to a 
vortex trap state.  This can be accomplished through a Raman scheme in 
which one of the two copropagating laser beams has a higher-order 
Laguerre-Gaussian 
mode profile.  Coefficients relating the beam waist, pulse area, 
and trap potentials for a complete transfer to the $m = 1$ vortex are 
calculated for a condensate in the non-interacting and strongly 
interacting regimes.

\end{abstract}


Based on the rapid experimental progress in the field of atomic 
Bose-Einstein condensation \cite{BECexpt,Andrews97,Myatt97}, the prospects look good that 
experimenters will soon be generating and observing vortices in these 
sytems.  Vortex solutions to the Gross-Pitaevski equation for the 
condensate wavefunction have already been found numerically 
\cite{Edwards96a,Dalfovo96}. It has been pointed out that unlike 
homogeneous Bose fluids such as $^4$He, there exist two distinct concepts 
of stability in a trapped Bose-Einstein condensate (BEC)\cite{Rokhsar}.
Although they are thermodynamically stable while the system is being 
rotated, this does not necessarily imply dynamical stability once the 
rotation ceases.  Thus, one might be interested in the dynamics of the 
vortex.  A necessary step is to start with a well-defined state; here 
we propose an optical method for coherently creating a vortex state 
from the ground state.

It is well known that the higher order Laguerre-Gaussian modes of a laser 
resonator possess orbital angular momentum \cite{Siegman86}.  (Note that this is possibly in 
addition to the angular momentum contributed by circular polarization of 
the light).  If this angular momentum can be coherently transferred to a 
BEC, the circular motion of the atoms should result in a vortex state.  
Since this involves removing photons from the Laguerre-Gaussian mode, we 
need to avoid subsequent spontaneous emission and heating of the 
condensate \cite{BabikerTwamley}.  One way to accomplish this is to use a Raman scheme in which 
the second beam is an ordinary Gaussian mode, and the atoms make a 
transition between hyperfine states along with the transfer of angular 
momentum.  To transfer all of the atoms from the trap ground state to the 
vortex state, we need to find the equivalent of a two-photon $\pi$ pulse.

We denote the condensate wavefunctions for hyperfine states $1$ and $2$ by
$\psi_1$ and $\psi_2$.  The trapping in current experiments is done 
magnetically, so we allow for different trap frequencies 
$\omega_1$, $\omega_2$ for the two atomic states. Since the vortex can 
already be described in two dimensions, we will treat the condensate as 
though it simply has a thickness $L$ in the $z$-direction.  The light beams 
both enter from the top (or bottom) and are copropagating.  An important 
assumption is that the condensate be optically thin in this dimension so 
that the light beams will affect all of the atoms. 
For the time being we neglect the collisions between atoms; their 
effect will be explained below. It is convenient to define
\bea
a_\bot	& =	& \sqrt{\frac{\hbar}{m}}(\omega_1 \omega_2)^{-\frac{1}{4}} \\
a_1	& =	& a_\bot \alpha^{\frac{1}{4}} \\
a_2	& =	& a_\bot \alpha^{-\frac{1}{4}} \\
\alpha & = & \frac{\omega_2}{\omega_1}
\eea
and use the scaling
\bea
\bar{r}	& =	& \frac{r}{a_\bot} \\
\bar{t}	& =	& \sqrt{\omega_1 \omega_2} \, t \\
\bar{\psi_j} (\bar{r})  & = & a_\bot^{\frac{3}{2}} \psi_j (r).
\eea
 Then, after adiabatically eliminating the third level from the Raman 
transition \cite{Javanainen95}, the modified Gross-Pitaevski equations are given in 
polar coordinates by 
\bea
i \donedtbar & = & \frac{1}{2} \left( -\bar{\nabla}^2 +	\frac{1}{\alpha} \bar{r}^2
\right)	\bar{\psi_1} + \bar{\Omega}	\bar{\psi_2} \\
i \dtwodtbar & = & \frac{1}{2} \left( -\bar{\nabla}^2 +	\alpha \bar{r}^2
\right)	\bar{\psi_2} + \bar{\Omega}^* \bar{\psi_1}.
\label{modGP}
\eea
Note that we use unit normalization,
\be
\int d^3 \, {\mathbf x} \left( |\psi_1|^2 + |\psi_2|^2 \right) = 1.
\ee
The two-photon Rabi frequency is 
\be
\bar{\Omega} = \frac{\Omega}{\sqrt{{\omega_1}{\omega_2}}}
\ee
and in the unscaled variables
\be
\Omega({\mathbf	x},t) = \frac{({\mathbf d}_1 \cdot {\mathbf E}_1({\mathbf 
x},t))^{*} 
{\mathbf d}_2 \cdot {\mathbf E}_2({\mathbf	x},t)}{\hbar^2 \Delta}
\label{Omunscale}
\ee
where $ {\mathbf d}_1, {\mathbf d}_2$ are the dipole matrix elements 
between states 1 and 3, 2 and 3 respectively, and ${\mathbf E}_1, {\mathbf 
E}_2$ are the near-resonant fields detuned by $\Delta$ from those 
transitions.  The fields have oppositely circular polarizations as 
appropriate to a $m_F$ to $m_{F+2}$ transition.  We will choose the field 
${\mathbf E}_1$ to be an ordinary Gaussian mode $({\mathrm TEM}_{00})$, while ${\mathbf E}_2$ will 
be the ${\mathrm TEM}_{11}$ Laguerre-Gaussian mode.  (In fact it makes no 
difference which of the two fields has which mode structure.) We approximate these 
modes by their values at $z=0$,
\bea
{\mathbf E}_1({\mathbf x},t)  & = & {\mathbf A}_1(t) \exp \left[ - r^2/w_1^2 \right] \\
{\mathbf E}_2({\mathbf x},t)  & = & {\mathbf A}_2(t) \frac{r}{w_2} \exp \left[ - 
r^2/w_2^2 \right] e^{i \phi}
\eea
Substituting into Eq. (\ref{Omunscale}) yields
\be
\Omega({\mathbf	x},t) =	\Omega_0(t)	\frac{r}{w_2}
\exp \left[	{-\frac{r^2}{w^2}} \right] e^{i \phi}
\ee
where
\be
\frac{1}{w^2} = \frac{1}{w_1^2} + \frac{1}{w_2^2}
\ee
and
\be
\Omega_0(t) = \frac{({\mathbf d}_1 \cdot {\mathbf A}_1(t))^{*} {\mathbf d}_2 \cdot {\mathbf 
A}_2(t)}{\hbar^2 \Delta}
\ee

We begin with all the atoms in hyperfine state 1 and the associated trap 
ground state.  The ground trap states of each hyperfine state are (we 
drop the bars on the scaled variables) 
\bea
\psi_1^{ (0) }(r) & = & \left( \frac{1}{\pi^2 \alpha} \right)^{\frac{1}{4}} \exp
\left[ - \frac{1}{2 \sqrt{\alpha}} r^2 \right] \\
\psi_2^{ (0) }(r) & = & \left( \frac{\alpha}{\pi^2} \right)^{\frac{1}{4}} \exp 
\left[ - \frac{\sqrt \alpha}{2} r^2 \right]
\eea
and the two vortex states rotating clockwise and counterclockwise are 
\bea
\psi_1^{(\pm 1)}(r,\phi) & = & \left( \frac{1}{\pi \alpha} \right)^{\frac{1}{2}} r \exp
\left[ - \frac{1}{2 \sqrt{\alpha}} r^2 \right] e^{\pm i \phi}\\
\psi_2^{(\pm 1)}(r,\phi) & = & \left( \frac{\alpha}{\pi} \right)^{\frac{1}{2}} \exp
\left[ - \frac{\sqrt \alpha}{2} r^2 \right] e^{\pm i \phi}
\eea
We expand the time-dependent condensate wavefunctions in this basis.
\bea
\psi_j({\mathbf x},t)  =  c_j^{(0)}(t) \psi_j^{(0)}({\mathbf x}) e^{-i 
\mu_j^{(0)} t} & + &
\nonumber \\ 
 c_j^{(+1)}(t) \psi_j^{(+1)}({\mathbf x}) e^{-i \mu_j^{(+1)} t} & + & 
 \nonumber \\
 c_j^{(-1)}(t) \psi_j^{(-1)}({\mathbf x}) e^{-i \mu_j^{(-1)} t} & \mbox{for} & \;j = 1,2 
\label{expand}
\eea
Here each $\mu_j^{(m)}$ is the chemical potential when all atoms 
are in the hyperfine state $j$ and the condensate has azimuthal 
angular momentum $m$.
We have neglected the higher terms of the series for two reasons. The Raman 
transition couples these higher trap states which are initially empty only 
to each other and therefore they
will always be unoccupied.  Second, when collisions between the atoms are 
included (see below), the differences in chemical potentials for higher 
states will no longer be on 
resonance so the transitions are weakly driven in any case.  Substituting 
into the Gross-Pitaevski equations (\ref{modGP}) and using the orthogonality of the 
$\psi_j^{(m)}$, we find the equations of motion
\bea
i \frac{dc_1^{(0)}}{dt} & = & g(t) c_2^{(-1)} e^{-i (\mu_2^{(-1)} - \mu_1^{(0)}) t} \\
i \frac{dc_2^{(-1)}}{dt} & = & g^*(t) c_1^{(0)} e^{i (\mu_2^{(-1)} - \mu_1^{(0)}) t}
\eea
where
\be
g(t) =  \bar{L} \int_0^{2 \pi} d \phi\int_0^\infty r \, dr \, 
\psi_1^{(0)*}({\mathbf x}) \psi_2^{(-1)}({\mathbf x}) \Omega({\mathbf x},t)
\label{gintegral}
\ee
These are exactly equivalent to the Bloch equations for a two-level 
system.  We note that an additional detuning $\mu_2^{(-1)} - \mu_1^{(0)}$ 
is required from the untrapped atom resonanant Raman transition; the 
energy this represents is 
imparted to the gas to produce the rotation. The effective Rabi frequency 
is given by
\be
g(t) = \bar{\Omega}_0(t) \bar{L} \alpha^{\frac{1}{4}}  \left( \frac{a_\bot^2}{w^2}  + \frac{\sqrt{\alpha}}{2}
 + \frac{1}{2 \sqrt{\alpha}} \right)^{-2}
\ee
As usual \cite{AllenEberly}, one may obtain a complete transfer to the upper 
level at time $t_f$ by 
choosing an on-resonance $\pi$-pulse,
\bea
g(t) & = & g_r(t) e^{i(\mu_2^{(-1)} - \mu_1^{(0)})} \\
g_r(t) & = & g_r(t)^* \\
\int_0^{t_f} g_r(t') \, dt' & = & \pi
\eea

We can expect that the technique will remain applicable when 
interactions between the atoms are included.  We replace the ground 
and vortex wavefunctions in Eq.  (\ref{expand}) by their Thomas-Fermi 
approximations in the strongly interacting regime,
\bea
\psi_1^{(0)} & = & \sqrt{\frac{2 \mu_1^{(0)} - \frac{r^2}{\alpha}}{u_{11}}} \\
\psi_2^{(\pm 1)} & = & \sqrt{\frac{2 \mu_2^{(\pm 1)} - \alpha r^2 - 
\frac{1}{r^2}}{u_{22}}} e^{\pm i \phi}
\eea
where
\be
u_{jj} = \frac{8 \pi N a_{jj}}{a_\bot}
\ee
is the scaled interaction potential between two atoms in hyperfine 
state $j$.  These functions then appear in the integral for the 
effective Rabi frequency Eq.  (\ref{gintegral}).  (Note that for the 
vortex, the Thomas-Fermi approximation includes the azimuthal kinetic 
energy but neglects the radial part.) The chemical potential of the 
ground state is found from normalization to be
\be
\mu_1^{(0)} = \sqrt{\frac{u_{11}}{2 \pi \alpha \bar{L}}}
\ee
To find the chemical potential of the vortex state, we 
must solve numerically the normalization equation
\be
\frac{u_{22}}{2 \pi\bar{L}} = \frac{\cosh \zeta}{\alpha \sinh^2 \zeta} - 
\ln \coth \frac{\zeta}{2}
\ee
and then set
\be
\mu_2^{(1)} = \coth \zeta.
\ee

It is desirable to make the effective Rabi frequency large, so that 
short pulses can be used and the dynamics of the condensate can be 
ignored while the vortex is being made.  To this end, we numerically 
evaluate the integral Eq.  (\ref{gintegral}) and determine its 
dependence on the waist $w_{1}$ and $w_{2}$ of each beam.  We choose 
parameters similar to those of the JILA experiment with overlapping 
condensates in two hyperfine states \cite{Myatt97}.  For a total of $N 
= 2\times 10^6$ condensed $^{87}$Rb atoms, mean of trap frequencies 
$\sqrt{\omega_{1} \omega_{2}} = 2 \pi \times 400$Hz, thickness $L = 
10^{-6}$m, $\alpha = \sqrt{2}$, and scattering lengths $a_{jj} = 
10^{-8}$m, we obtain the ground and vortex states shown in Fig. 
\ref{waveffig}.  
(The mean trap width is $a_{\bot} = 540$nm.)  The required detuning 
$(\mu_{2}^{(0)} - \mu_{1}^{(-1)})/ \hbar = 39.4$kHz is mostly due to 
the difference in trap potentials in this example.  In Fig. 
\ref{overlfig} 
we show the dependence of $g/\Omega_{0}$ on the beam waists.  We note 
that there is a maximum as a function of $w_{2}$ when $w_{1}$ is held 
fixed, for a value of $w_{2}$ on the order several $a_{\bot}$.  It is 
advantageous to choose $w_{1}$ as large as possible; for the case of a 
plane wave $\mathbf A_{1}$, the optimal waist size for the 
Laguerre-Gaussian beam is $w_{2} = 19.3a_{\bot}$.


The resulting vortex state has a zero condensate density at its 
center.  However, current techniques are probably not capable of 
imaging this small core in the atomic density directly.  In liquid 
$^{4}$He, a negative ion can be trapped at the vortex core, and after 
accelerating the ion out of the fluid it can be imaged on a phosphor 
screen \cite{Williams80}.  It might be possible to adapt this 
technique to the atomic BEC.  An alternative way of observing the 
vortex is to allow it to interfere with another (non-vortex) 
condensate, and image the fringes as in \cite{Andrews97}.  A similar 
technique has been used in nonlinear optics \cite{Swartzlander93}.  
The $2 \pi$ phase shift around the vortex will show up as two 
interference fringes merging into one.

We have shown how a vortex state can be created in a trapped Bose-Einstein 
condensate.  The transfer to the new state is completely coherent, as it 
involves only stimulated optical transitions.  It could also be applied to 
create higher vortex states with $|m|>1$. To investigate persistent 
currents, one may transfer the atoms to the vortex while a repulsive 
potential is placed at the center; a simple way to do this is to detune the 
Gaussian laser mode above the single-photon resonance and leave that 
field on after the Laguerre-Gaussian pulse has ended.

\section*{Acknowledgments}
This research was funded by the Marsden Fund of the Royal Society of 
New Zealand, and University of Auckland Research Fund.  We would like 
to thank M.\ Andrews, D.\ Rokhsar, M.\ Levenson, T.\ Wong and M.\  
Olsen for helpful discussions.

\begin{figure}
\caption{Thomas-Fermi wavefunctions of the initial and final 
condensate, appropriate for strongly interacting atoms, for the 
parameters given in text.
{\em Solid curve}:  Ground trap state, hyperfine state 1 {\em Dashed curve}:
Vortex trap state, hyperfine state 2. Note this function has a small 
core at which the density goes to zero.}
\label{waveffig}
\end{figure}

\begin{figure}
\caption{Overlap integral $g/\Omega_{0}$ as a function of 
Laguerre-Gaussian beam waist $w_{2}$ for $w_{1} = \infty$ (solid 
curve), $w_{1} = 50 a_{\bot}$ (long dash) and $w_{1} = 20 a_{\bot}$ (short dash).}
\label{overlfig}
\end{figure}

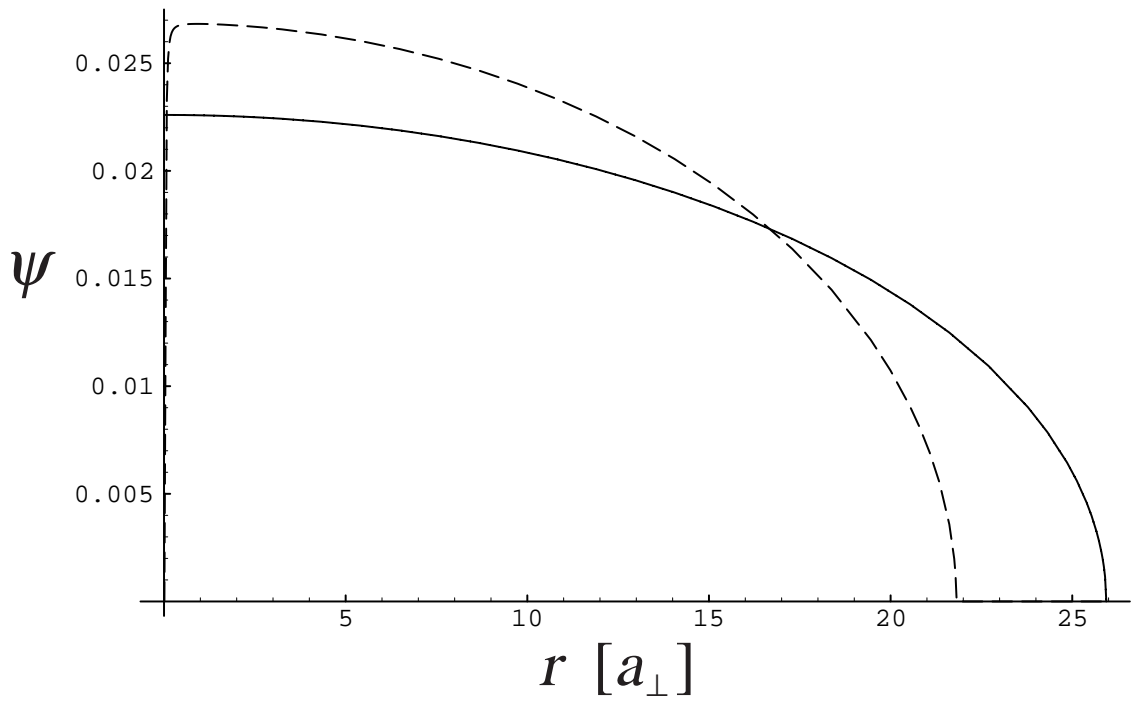
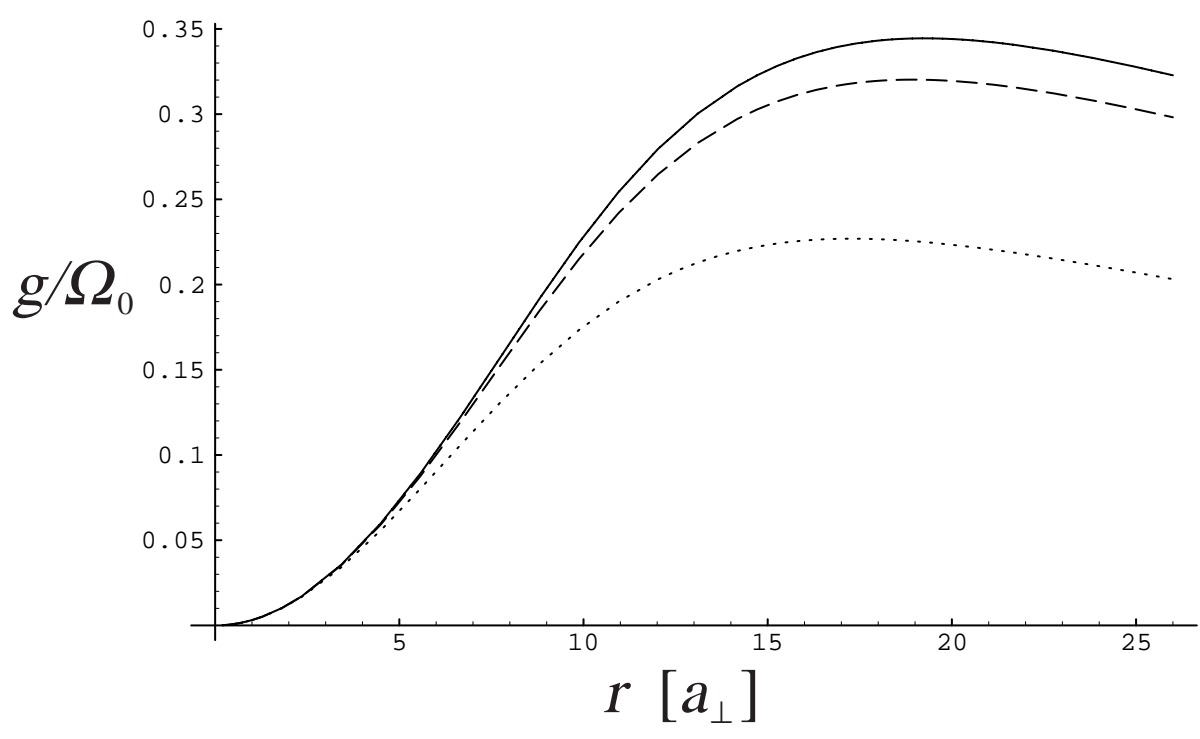
\end{document}